\documentclass[aps,pre,reprint]{revtex4-1}
\usepackage{graphicx,bm}

\newcommand{\etal}{\emph{et~al. }}

\usepackage[normalem]{ulem}

\begin{document}

\title{Dimensional transition in rotating turbulence}

\author{E. Deusebio$^{1}$, G. Boffetta$^{2}$, E. Lindborg$^{1}$, S. Musacchio$^{3}$}
\affiliation{
$^1$Linn\'e Flow Centre, Department of Mechanics, 
Royal Institute of Technology, 10044 Stockholm, Sweden\\
$^2$Dipartimento di Fisica Generale and INFN, 
Universit\`a di Torino, via P.Giuria 1, 10125 Torino, Italy\\
$^3$CNRS, Lab. J.A. Dieudonn\'e UMR 6621,
Parc Valrose, 06108 Nice, France}
\date{\today}

\begin{abstract}
In this work we investigate, by means of direct numerical
hyperviscous simulations, how rotation affects the bi-dimensionalization
of a turbulent flow.
We study a thin layer of fluid, forced by a two-dimensional
forcing, within the framework of the "split cascade" in which
the injected energy flows both to small scales (generating the
direct cascade) and to large scale (to form the inverse cascade).
It is shown that rotation reinforces the inverse cascade
at the expense of the direct one, thus promoting
bi-dimensionalization of the flow. This is achieved 
by a suppression of the enstrophy production at large scales.
Nonetheless, we find that, in the range of rotation rates investigated, 
increasing the vertical size of the computational domain causes a 
reduction of the flux of the inverse cascade. 
Our results suggest that, even in rotating flows, the inverse cascade   
may eventually disappear when the vertical scale is sufficiently large 
with respect to the forcing scale.
We also study how the split cascade and confinement influence the breaking of symmetry induced by rotation.
\end{abstract}

\maketitle

\section{Introduction}
\label{sec:1}

All turbulent processes are characterized by a transfer of kinetic energy
among scales. However, depending on the dimension of the space, completely
different phenomenology appear. In three dimensions (3D), energy is transferred
in a direct cascade towards small scales
\cite{MY71,Frisch95}, whereas in two dimensions (2D) energy
undergoes an inverse cascade towards large scales \cite{Kraichnan67,BE12}.
Numerical simulations have confirmed this picture in both 
3D \cite{IGK09} and 2D turbulence \cite{BM10,VL11} but they have 
also shown that the two processes are not mutual exclusive and 
the coexistence of both a downscale and upscale energy transfer has been
observed \citep{SCW96,MP09,CMV10,VDL11,DVL13}.

Previous numerical investigations \cite{SCW96,CMV10} 
studied the cross-over between 2D and 3D
dynamics by considering triply-periodic domains where one dimension was
increasingly contracted. As the ratio $S=L_z/L_f$ between 
the geometrical dimension of the compactified direction, $L_{z}$, and 
the length scale of the forcing, $L_f$, decreased below $\sim 0.5$, 
a mixture of 2D and 3D dynamics was
observed with a coexistence of a forward and an inverse energy cascade. 

In a similar way, also rotation favors a two-dimensionalization of the flow,
generally supported by the \emph{Taylor-Proudman} theorem \cite{Taylor23}.
Experiments of decaying turbulence in rotating tanks show the growth of length
scales aligned with the rotation axis, giving evidence that turbulence
developing in rotating systems is highly anisotropic. Both linear
\cite{DSD06,SDD08} and non-linear \cite{CMG97,SW99} mechanisms have been
advocated to explain the observed growth. Numerical simulations of rotating
flows have also indicated a general trend towards two-dimensionalization
\cite{SCW96,SW99} and
some studies \cite{MP09} have reported a split of the energy cascade into a
downscale and an upscale process.  Other studies \cite{CCEH05} have found
that, as the rotation rate $\Omega$ increases, the flows exhibit a dynamics
very similar to what is found in a 2D system, with the vertical velocity
behaving as a passive scalar \cite{EM98}.  Bourouiba \& Bartello \cite{BB07}
reported a strong transfer of energy from 3D modes to 2D modes at intermediate
rotation rates, whereas at large rotations a decoupling of the dynamics was
observed. 
The localness of such energy transfer has been investigated in \cite{BSW12}. 

However, it is not obvious how, starting from a 3D flow fields, the actual
two-dimensionalization would occur. The effect of rotation in a turbulent
system has recently been a matter of debate.  
In the strongly rotating limit, energy
transfer among scales is supposedly dominated by resonance triad interactions
between inertial waves \cite{W93}. Even though such interactions can move
energy towards the 2D plane \cite{SW99}, triadic resonance cannot transfer
energy into/out of the $k_z=0$ plane, which in fact constitutes a closed
resonant set. If only interactions among resonant triads are allowed, 2D modes
should thus be dynamically decoupled from 3D modes \cite{EM98,CCEH05}.
Sometimes, it is advocated that nearly-resonant interactions \cite{SW99}, for
which the resonance condition is satisfied only to a certain degree, and
higher-order resonance interactions \cite{Newell69} can explain the transfer of
energy into the $k_z=0$ plane.  However, as the flow tends towards
two-dimensionalization, the advective time scale, $t_a \sim U/ l_h$, and the
inertial wave time scale, $t_w \sim l_z / \Omega l_h $, approach each other and
transfer of energy may not only be determined by resonant triads.  Cambon \etal
\cite{CRG04} suggested that in the framework of wave turbulence theory,
two-dimensionalization cannot be achieved for unbounded domains, even in the
limit of infinite rotations, and a coupling between 2D and 3D modes always
exists. 

The interest in three-dimensional dynamics in rotating turbulence has
recently been revived also by experiments carried out on a large-scale
\emph{Coriolis} platform. Moisy \etal \cite{MMRS11}
have shown that flow fields originating from grid-generating decaying
turbulence remain highly 3D, even at large rotation rates.  
Vertical velocity was not behaving as a passive scalar and its
coupling with the large-scale flow was suggested to trigger shear instabilities.
Thus, it is not clear whether rotating flows would approach a 2D-like dynamics
and whether 2D modes would ever dynamically decouple from the 3D flow field. 

The effects of rotations are even more complex in the case in which 
the flow is confined between two parallel walls perpendicular 
to the rotation axis. Numerical simulations have shown reflections 
of inertial waves on the walls 
and a transition to an almost two-dimensional state\cite{GL99}. 
The wave turbulence regime which develops in the limit of strong rotation 
in a flow confined in the direction of the rotation axis has been 
recently studied \cite{S14}. 

Another remarkable feature of rotating turbulent flows 
is the breaking of symmetry between cyclonic and anticyclonic vortices. 
The predominance of cyclonic vortices 
(\emph{i.e.} those which spin in the same direction of ${\bm \Omega}$)
has been observed both in experiments
\cite{HBG82,HV93,MMR05,SDD08,MMRS11} 
and numerical simulations 
\cite{bartello_metais_lesieur_1994,BB07,BCLGC08}.
Several explanations have been proposed to explain this phenomenon.  
In particular, it has been shown that the cyclones and anticyclones 
have different stability properties \cite{bartello_metais_lesieur_1994} 
and different probabilities to be generated at finite Rossby number \cite{SD08}. 
Moreover, the correlations between the strain tensor and the vorticity 
in isotropic turbulence can be responsible for 
the development of a positive skewness of vertical vorticity
when the flow is suddenly subjected to rotation \cite{GF01}.  
Interestingly, a similar asymmetry has been observed also in the tropopause
\cite{HC05} and in the stratosphere \cite{LC01}, even though the physical mechanisms acting in the atmosphere are
more complex than in idealized rotating turbulence.

Previous works have investigated the dependence of the asymmetry on intensity
of rotation.  In particular it has been shown that the skewness of vertical
vorticity has a non monotonic behavior as a function of the Rossby number \cite{BB07}.  The asymmetry disappear both in the limit of strong and weak
rotation and it attain its maximum for intermediate rotation rates \cite{BB07}.
On the contrary it is still unclear how this asymmetry depends on the aspect
ratio $S$ of the flow.   

In this paper we investigate how the combined effect of rotation
and a periodic confinement affects the turbulent dynamics. 
We consider the case of intermediate rotation intensities, 
such that the Coriolis forces are neither too weak to be neglected 
nor too strong to overwhelm non-linear interactions. 
We mainly focus on the transfer of energy and we
investigate whether statistically steady regimes, where all the injected 
energy is transferred towards small scales, can be achieved in rotating 
confined flows.  
We also investigate how the parity symmetry breaking on the horizontal
plane induced by rotation is affected by the rotation 
and by the aspect ratio of the flow. 

The paper is organized as follows: Section~\ref{sec:2} discusses the numerical 
code and the parameters used in the simulations; Section~\ref{sec:3} is 
devoted to the effect of rotation and confinement on the upscale and downscale 
cascades. Section~\ref{sec:4} studies the asymmetry in the vorticity
field induced by rotation and confinement and Section~\ref{sec:5} 
is devoted to conclusions.

\section{Numerical simulations of rotating turbulence}
\label{sec:2}
We consider the 3D Navier-Stokes equations for an incompressible 
velocity field ${\bm u}({\bm x},t)$ (\emph{i.e.} ${\bm \nabla}\cdot {\bm u}=0$) in a rotating frame of reference,
\begin{equation}
{\partial {\bm u} \over \partial t}+ {\bm u} \cdot {\bm \nabla} {\bm u} +
2 {\bm \Omega} \times {\bm u}= - {\bm \nabla}p + \nu \nabla^2 {\bm u}
+ {\bm f},
\label{eq:2.1}
\end{equation}
where the constant density has been absorbed into the pressure $p$, 
$\nu$ is the kinematic viscosity and ${\bm \Omega}=\Omega {\boldsymbol z}$
represents the constant rotation along the vertical direction aligned with the  unit vector $\boldsymbol z$
(we remark that here "vertical" is used only in analogy with the 
direction of confinement in experiments as gravity does not explicitly
appear in equation \ref{eq:2.1}).
The forcing field ${\bm f}({\bm x},t)$ is a stochastic Gaussian, white
in time noise, active only on the horizontal components $u_x$, $u_y$
of the velocity and depends on the horizontal components $x$, $y$ only.
The forcing is localized in Fourier space in a narrow band of wave
numbers around $k_f=2 \pi/L_f$ and injects energy into the system at a fixed
rate $\varepsilon_I$ \cite{lindborg_brethouwer_2007}.

Direct numerical simulations of (\ref{eq:2.1}) are performed by means of
a $2/3$-dealiased, parallel,
 pseudo-spectral code in a triply periodic domain with
various aspect ratios, $r=L_x/L_z$, and rotations. 
Simulations are carried out with uniform
grid spacing at resolution $N_x=N_y=r N_z$ with fixed $N_x=512$, $L_x=2\pi$  and 
forcing wavenumber $k_f=8$.
The linear rotation and viscous terms are integrated using an exact 
factor technique, thus removing them from the explicit time integration scheme 
which uses a second-order Runge-Kutta 
\citep{canuto_hussaini_youssuff_quarteroni_zang_2006}.
The viscous terms in (\ref{eq:2.1}) is replaced by a second-order hyper-viscosity (Laplacian square) to increase the extension of the inertial
range. The hyper-viscosity coefficient $\nu$ is chosen such that $k_{max} \eta \approx 1.3$, with $\eta= \left( \nu^3 / \varepsilon_I \right)^{1/10}$, similar to what is required in a resolved DNS.

Previous simulations in the absence of rotation \cite{SCW96,CMV10} 
showed that the ratio between the vertical scale and the forcing 
scale $S=L_z/L_f$ controls the relative amount of energy that flows
to large scales. In particular, it has been shown that for
$S \ge 1/2$ the inverse energy flux vanishes and the turbulent flow
recovers a 3D phenomenology \cite{CMV10}.
A second dimensionless parameter in our simulations is provided by
the rotation rate which defines a rotation number when made
dimensionless with the characteristic time at the forcing scale,
$R \equiv \Omega/(k_f^2 \varepsilon_I)^{1/3}$ ($R$ is essentially the
inverse of the Rossby number defined as in \cite{SW99}).
In table~\ref{table1} we show the range of parameters in the $(S,R)$
plane in which we performed our simulations.

\begin{table}[t!]
\begin{tabular}{c||c|c|c|c|c|c|c|c|c|c}
$R \backslash S$ & $0.125$ & $0.188$ & $0.250$ & $0.375$ & $0.50$ & $0.75$ & $1.0$ & $2.0$ & $4.0$ & $8.0$ \\ \hline
$0.0$  & $0.24$ & $0.71$ & $0.91$ & $0.97$ & $1$ & $1$ & $1$ & $1$ & - & - \\
$0.5$ & $0.22$ & $0.66$ & $0.81$ & $0.95$ & $1$ & - & $1$ & $1$ & - & - \\
$0.75$ & $0.18$ & $0.61$ & $0.73$ & $0.89$ & $0.89$ & $0.94$ & $1$ & $1$ & - & - \\
$1.0$ & $0.17$ & $0.45$ & $0.64$ & $0.76$ & $0.77$ & $0.82$ & $0.86$ & $0.87$ & $0.95$ & $1$ \\
$1.5$ & $0.14$ & $0.31$ & $0.42$ & $0.46$ & $0.49$ & $0.52$ & $0.52$ & $0.58$ & $0.66$ & $0.89$ \\
$5.0$ & - & - & - & - & - & - & $0.02$ & $0.02$ & $0.08$ & $0.14$ \\
$10.0$ & - & - & - & - & - & - & $0.01$ & $0.01$ & $0.01$ & $0.02$ 
\end{tabular}
\caption{Parameter space of the simulations. Each number represents the 
fraction of energy dissipated at small scales,
$\varepsilon_{\nu}/\varepsilon_I$, as a function of 
$R=\Omega/(k_f^2 \varepsilon_I)^{1/3}$ and $S=L_z/L_f$. 
The value $1$ means that the measured value is compatible with $\varepsilon_{inv}=0$ and 
therefore a case with pure 3D phenomenology. }
\label{table1}
\end{table}

We study the transition from 2D to 3D turbulence by looking at the evolution of
the mean kinetic energy $E(t)=\langle |{\bm u}({\bm x},t)|^2 \rangle/2$ for a
flow starting from low-amplitude random noise. For a 3D turbulent flow, after
an initial
transient, the mean kinetic energy attains a statistically steady state and the
viscous energy dissipation 
$\varepsilon_{\nu}=\nu \langle ({\bm \nabla} {\bm u})^2 \rangle$ 
balances the input $\varepsilon_I$.  In presence of an inverse cascade,
a part of the injected energy, $\varepsilon_{inv}$, flows 
to large scales and is not dissipated by
viscosity.  Energy conservation requires that
$\varepsilon_I=\varepsilon_{inv}+\varepsilon_{\nu}$.  
Therefore, a signature of an inverse cascade is a linear growth 
of the mean kinetic energy,
$\varepsilon_{inv}=\mathrm{d}E/\mathrm{d}t$, 
and consequently a viscous dissipation
$\varepsilon_{\nu}<\varepsilon_I$.  
We remark that in the presence of an inverse cascade,
the flow does not attain a statistically stationary state, as kinetic energy
continues to grow and larger and larger scales are generated. This process is
nonetheless very slow as the characteristic time grows with the scale $r$
following Kolmogorov scaling $r^{2/3}$. Therefore, even in this case we observe
a quasi-stationary regime in which, on the time scale of the largest active
modes, small scale statistics can be averaged.

\section{Dimensional transition and the effect of rotation}
\label{sec:3}

\begin{figure}[htb!]
\includegraphics[width=1.\columnwidth]{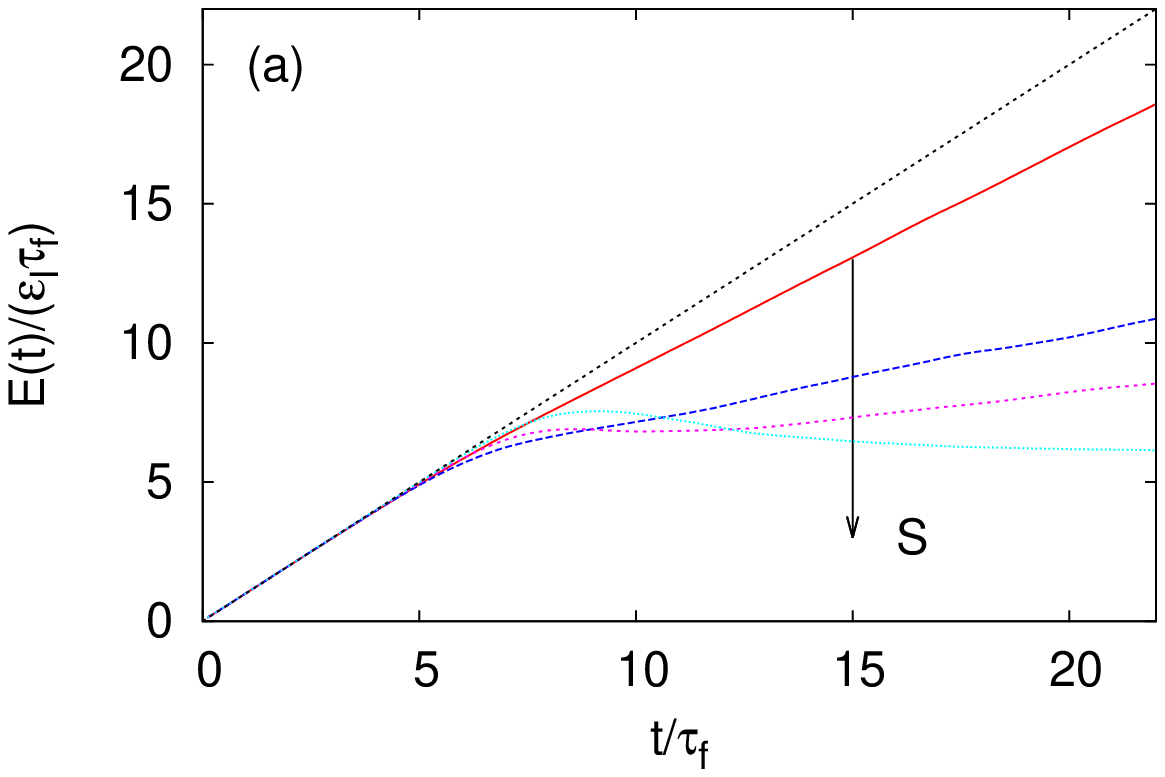} \\
\includegraphics[width=1.\columnwidth]{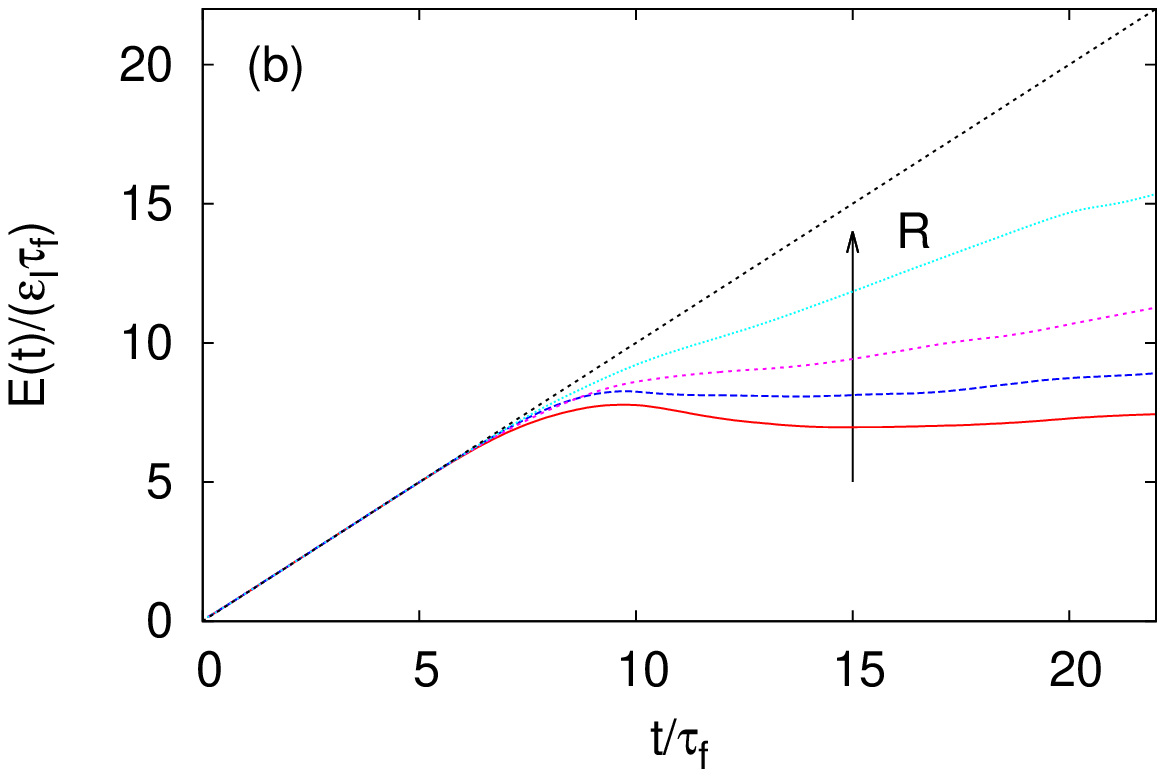}
\caption{(a) Time evolution of mean kinetic energy for runs at $R=0$ 
and different values of $S=0.125,0.1875,0.25,0.5$ (from top to bottom).
(b) Time evolution of mean kinetic energy for runs at $S=1/2$ and 
different values of $R=0.5,0.75,1.0,1.5$ (from bottom to top).
Time is normalized with the characteristic forcing time 
$\tau_{f}=(k^2 \varepsilon_I)^{-1/3}$. 
The dotted straight line in both plots represents the energy 
injection rate.}
\label{fig1}
\end{figure}

Figure~\ref{fig1}a shows the evolution of the mean kinetic energy, $E(t)$, for
different values of the aspect ratio $S$ in the absence of rotation ($R=0$).
After an initial transient (of duration independent on $S$) in which turbulence
develops, we observe a linear growth of the kinetic energy at a rate which is
smaller than the input. The linear growth rate, which defines $\varepsilon_{inv}$,
diminishes as the aspect ratio $S$ increases and eventually vanishes for $S
\simeq 1/2$ as shown in previous work \cite{SCW96,CMV10}. The value of
$\varepsilon_{inv}$ has been estimated with a linear least-square fit of $E(t)$. We remark that the threshold, $S=1/2$, is not expected to be an universal
value, as it depends on the particular forcing and also on the precise
definition of $k_f$. Indeed it has been shown that different forcing schemes
lead to different values of $S$ at which the inverse flux vanishes
\cite{SCW96}.

Figure~\ref{fig1}b illustrates the effect of rotation on the split of the
energy cascade. At fixed $S=1/2$, by increasing $R$ above zero, an inverse
cascade is recovered (for $R>0.5$ approximately) 
with a flux which increases with $R$.  This scenario is
reproduced at all values of $S$ which we have investigated: for sufficiently
large values of rotation an inverse cascade is recovered, as shown in
table~\ref{table1}.

\begin{figure}[htb!]
\includegraphics[width=1.\columnwidth]{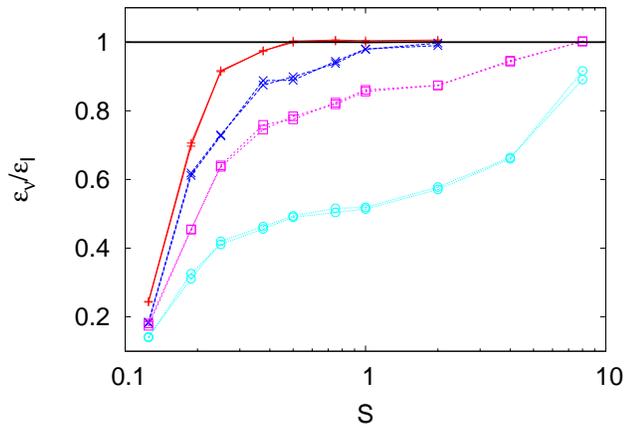}
\caption{Ratio of small-scale energy dissipation $\varepsilon_{\nu}$
to the energy input $\varepsilon_I$ as a function of the aspect ratio
$S$ for different rotation numbers, $R=0,0.75,1,1.5$ (from
top to bottom). For each $R$ we plot two lines, one given directly by
$\varepsilon_{\nu}/\varepsilon_{I}$, the second by $\varepsilon_{inv}/\varepsilon_I$
in order to have an estimation of the statistical errors.}
\label{fig2}
\end{figure}

We remark that, because we are in quasi-stationary conditions, we can measure
the inverse cascade flux also as
$\varepsilon_{inv}=\varepsilon_I-\varepsilon_{\nu}$.  
The two values obtained differ by a few percent, because of the errors in the
linear fit of the energy growth and the statistical uncertainty 
of $\varepsilon_{\nu}$. In the following we use this difference as a measure of the errors in the calculation of $\varepsilon_I$ and $\varepsilon_\nu$.
In table~\ref{table1} the values of $\varepsilon_\nu$ in the $S-R$ parameter space are reported. 

The results obtained by this procedure are shown in Fig.~\ref{fig2} where the
ratio $\varepsilon_{\nu}/\varepsilon_I$ is plotted as a function of $S$ 
for different values of $R$. For the case without rotation, $R=0$, 
we see that for $S \geq 0.5$ the inverse cascade vanishes, as shown in 
Fig.~\ref{fig1}a.
Figure~\ref{fig2} shows that an increase of the rotation rate $R$ gives a
transition to a pure 3D regime at larger values of $S$.  The runs with strong
rotation, $R \ge 1.5$ (see table~\ref{table1}), show no dimensional transition
in $S$.  Nonetheless, the small-scale energy dissipation fraction always
increases with $S$ and no saturation at a value of
$\varepsilon_{\nu}/\varepsilon_I<1$ is evident. This observation suggests that
dimensional transition is always present in this system, even for strong
rotation rates, for large enough $S$.  Of course, this possibility could be
confirmed only by increasing $S$ but this would require simulations at larger
resolutions (at $S=8$ we already have $N_z=N_x$).

\begin{figure}[htb!]
\includegraphics[width=1.\columnwidth]{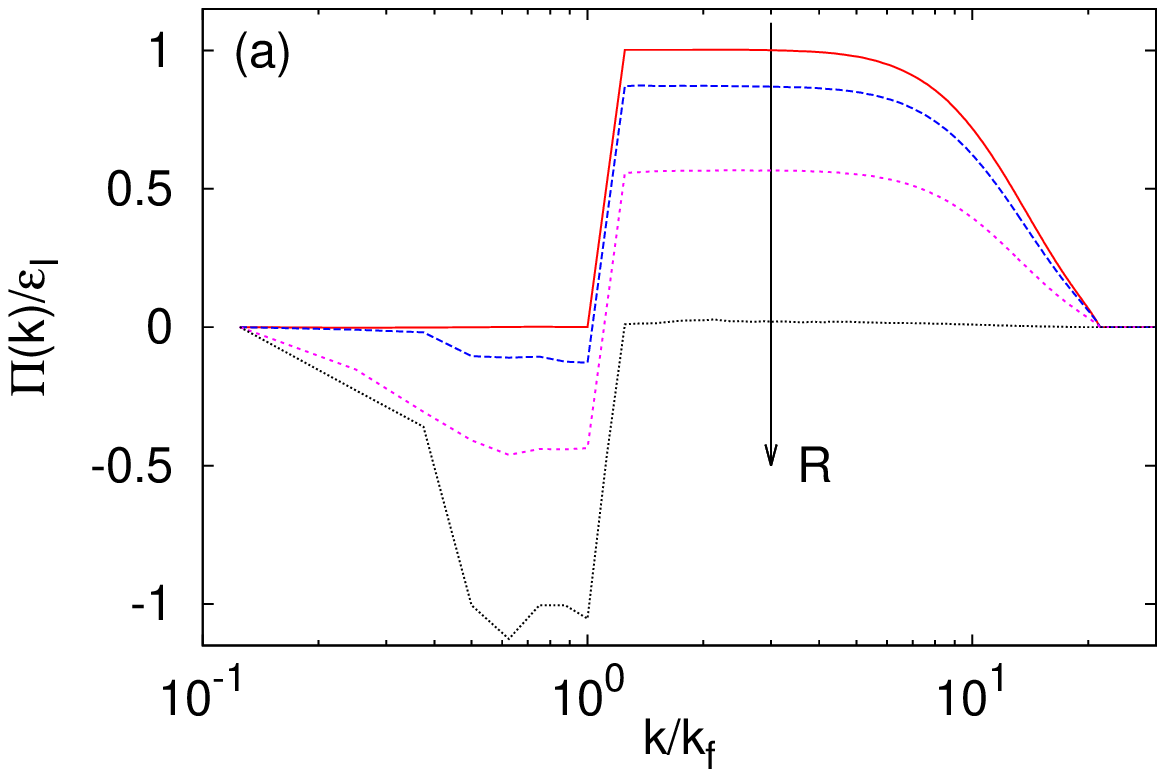} \\
\includegraphics[width=1.\columnwidth]{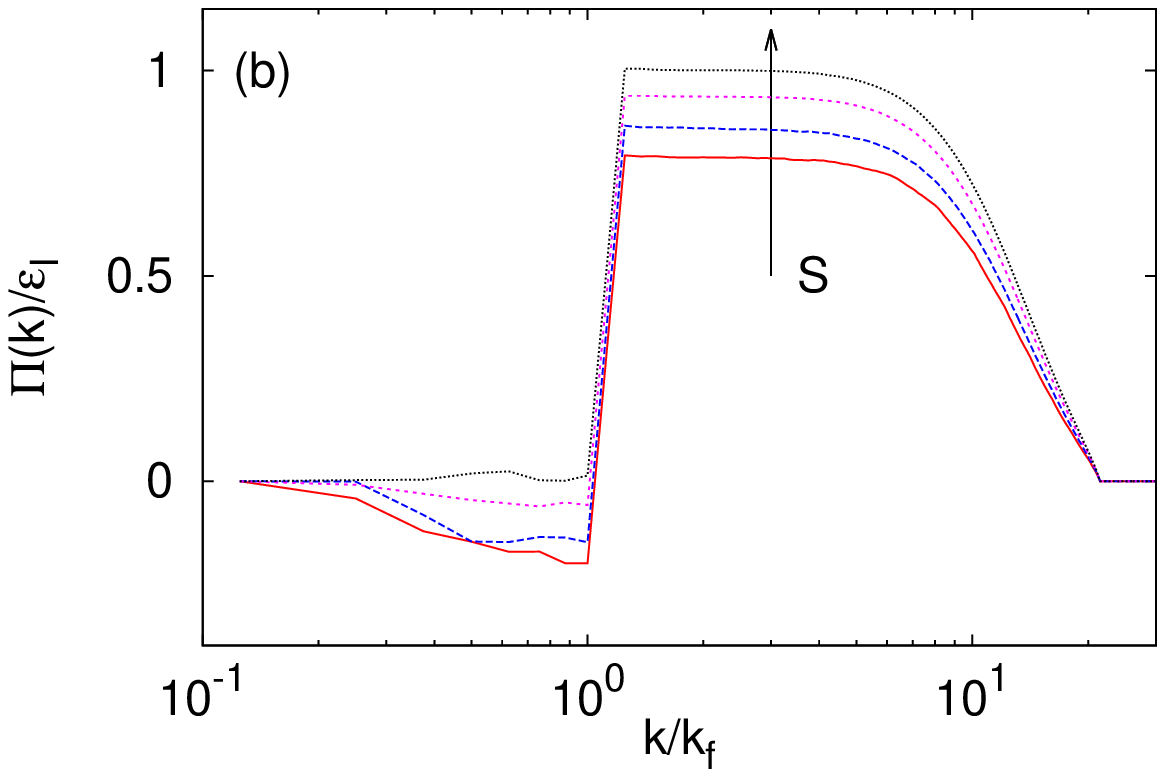}
\caption{(a) Spectral flux of kinetic energy for different 
values of $R=0,1,1.5,5$ (following the arrow) at fixed
$S=2$.
(b) Spectral flux of kinetic energy for different 
values of aspect ratio $S=0.5,1,4,8$ (following the arrow) at fixed
$R=1$. All fluxes are normalized with energy input $\varepsilon_I$.
}
\label{fig3}
\end{figure}

\begin{figure}[htb!]
\includegraphics[width=1.\columnwidth]{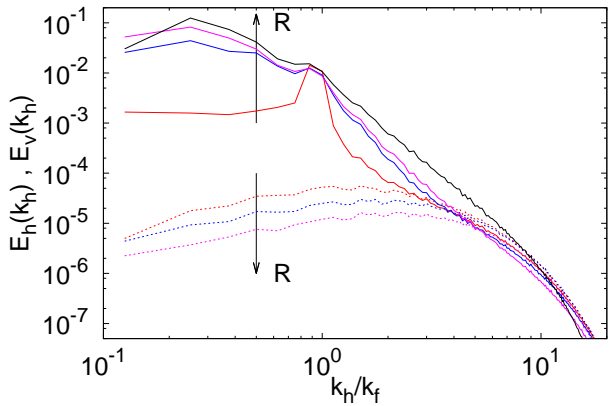}
\caption{
Kinetic energy spectra of the horizontal components $E_h(k_h)$ (solid
lines) and of the vertical component $E_v(k_h)$ (dotted lines) 
as a function of the horizontal wavenumber $k_h$ with $k_z=0$
for different rotations 
at $S=2$. Non-rotating case $R=0$ (red), $R=1.0$ (blue), $R=1.5$ 
(pink) and $R=5$ (black). Rotation increases from bottom to top for
continuous lines and decreases from top to bottom for dotted lines
(the vertical spectrum at $R=5$ is below $10^{-7}$ and is not
shown in the plot).
}
\label{fig4}
\end{figure}

The fact that part of the energy is not dissipated at small scales is not
sufficient to guarantee the presence of an inverse cascade as observed in 2D
turbulence. Indeed, an inverse cascade means a transfer of energy to
larger scales which can be observed only by looking at the flux of energy at
different wave numbers. 
We have therefore computed the spectral flux of kinetic energy for different
values of $R$ at given aspect ratio $S=2$ (shown in Fig.~\ref{fig3}a) and for
different $S$ at fixed rotation $R=1$ (see Fig.~\ref{fig3}b).  For $R=0$,
because $S=2$ is above the dimensional transition, we have a vanishing inverse
flux to wave numbers $k<k_f$ and a positive energy fluxes (equal to the energy
input) to wave numbers $k>k_f$, typical of a three-dimensional scenario with a
direct cascade to small scales.  By increasing $R$ the flux to small scales
reduces (in agreement with Fig.~\ref{fig2}), nonetheless, in all cases we
observe a clear plateau for wave numbers $k>k_f$.  It is interesting to observe
that for strong rotation rate (run at $S=2$ and $R=5$), the direct flux almost
vanishes, as predicted in a pure 2D scenario.  At small wave numbers, $k<k_f$,
and for $R>0$, we observe the development of an inverse cascade produced by
rotation. In this range of wave numbers the fluxes are more noisy but
nonetheless they are negative, which is the signature of an energy cascade
towards large scales.  We recall that the range of scales available for the
inverse cascade is quite small, since $k_f=8$.  Figure~\ref{fig3}b shows the
fluxes for different aspect ratios at fixed $R=1$. Again, for all the
simulations we have a clear plateau for $k>k_f$, at a value $\varepsilon_{\nu}$
which increases with $S$ (see Fig.~\ref{fig2}). The inverse cascade to
wavenumber $k<k_f$ is suppressed by increasing $S$ and eventually vanishes for
$S=8$.

Figure~\ref{fig4} shows the kinetic energy spectra corresponding to 
different simulations at $S=2$. The spectra of both the horizontal components, 
$E_h(k_h)$, and the vertical component, $E_v(k_h)$, of the energy are
plotted as a function of the horizontal wavenumber $k_h=\sqrt{k_x^2+k_y^2}$
with $k_z=0$.
The horizontal spectra, $E_h(k_h)$, display a clear peak around the 
forcing wavenumber $k_f$ and a narrow power-law scaling at larger 
wavenumber. For $k<k_f$ the energy spectra $E_h(k_h)$ strongly 
depends on rotation: by increasing $R$ we observe that more 
energy is present in the large scale modes. Because of the limited
scale separation between the forcing and the box scale ($k_f=8$)
we are not able to observe a clear Kolmogorov-like spectrum 
in the range of wavenumbers $k<k_f$.

In the range of small wavenumbers, the spectrum of the vertical
component, $E_v(k_h)$, is strongly suppressed with respect to the 
horizontal ones, $E_h(k_h)$, becoming even smaller as 
$R$ increases. At large wavenumber, the vertical spectra
become comparable (or even larger) than the horizontal spectra and
the $R$ dependence becomes weaker.

The fact that both the confinement and the rotation favor 
the development of the inverse cascade  
leads to an interesting consideration. 
Different flows can have the same ratio between 
inverse and direct energy fluxes for different values of 
$S$ and $R$, as it evident from Figure~\ref{fig2}. 
In particular a non-rotating thin layer can have the same flux ratio 
as a thick rotating one. 
In order to understand the similarities and differences between 
these two cases we need to investigate the mechanisms which are responsible 
for the transfer of energy towards large scales. 

\begin{figure}[htb!]
\includegraphics[width=1.\columnwidth]{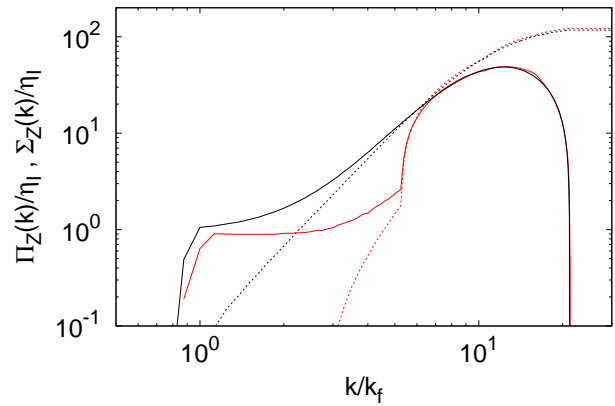}
\caption{
Spectral flux of enstrophy $\Pi_Z(k)$ (solid lines) and
enstrophy production $\Sigma_Z(k)$ (dashed line) for two flow configuration:
$R=0, S=0.188$ (lower, red lines) with $k_z=42.7$  and $R=1.5, S=4$ 
(upper, black lines) with $k_z=2$.
All quantities are normalized with the enstrophy input $\eta_I$.}
\label{fig5}
\end{figure}

The main difference between 3D and 2D Navier--Stokes equations, written for the
vorticity field, is the absence of the vortex stretching term, ${\bm \omega}
\cdot {\bm \nabla} {\bm u}$, in the latter. As a consequence, the enstrophy,
\emph{i.e.} mean square vorticity, is conserved in the inviscid limit in 2D
flows.  In the forced-dissipated case, the presence of two positive-defined
inviscid invariants (energy and enstrophy) causes the reversal of the direction
of the energy cascade with respect to the 3D case, and the simultaneous
development of a direct enstrophy cascade \cite{Kraichnan67,BE12}.  It is
therefore natural to investigate if a similar phenomenology can be observed
also in thin fluid layers.  In particular, we conjecture that the development
of the inverse cascade can be accompanied by a partial suppression of the
enstrophy production induced either by the confinement or by the rotation.  

To address this issue we computed the spectral flux of enstrophy $\Pi_Z(k)$ and
enstrophy production $\Sigma_Z(k)$ defined as:
\begin{equation}
\Pi_Z(k) = \int_{| \boldsymbol q | \le k}  
\widehat{ {\boldsymbol u} \cdot {\nabla} {\boldsymbol \omega} } \left( \boldsymbol q \right) \widehat{ \boldsymbol{\omega}}^*  \left( \boldsymbol q \right) \mathrm{d}{\boldsymbol{q} } ,
\label{eq:3.1}
\end{equation}

\begin{equation}
\Sigma_Z(k) = \int_{| \boldsymbol q | \le k}  
\widehat{ {\boldsymbol \omega } \cdot {\nabla} {\boldsymbol u} } \left( \boldsymbol q \right) \widehat{ \boldsymbol{\omega}}^*  \left( \boldsymbol q \right) \mathrm{d}{\boldsymbol{q} } ,
\label{eq:3.2}
\end{equation}
where $\widehat{\, \cdot \, }$ represents the three-dimensional 
Fourier transform. 
In Figure~\ref{fig5} we show both quantities for a non-rotating thin layer
$(R=0, S=0.188)$ and a rotating thick layer $(R=1.5, S=4)$.  As shown in
table~\ref{table1}, these two flows have approximatively the same ratio 
between the fluxes of the inverse and direct energy cascade.

\begin{figure*}[htb!]
\includegraphics[width=1.\columnwidth]{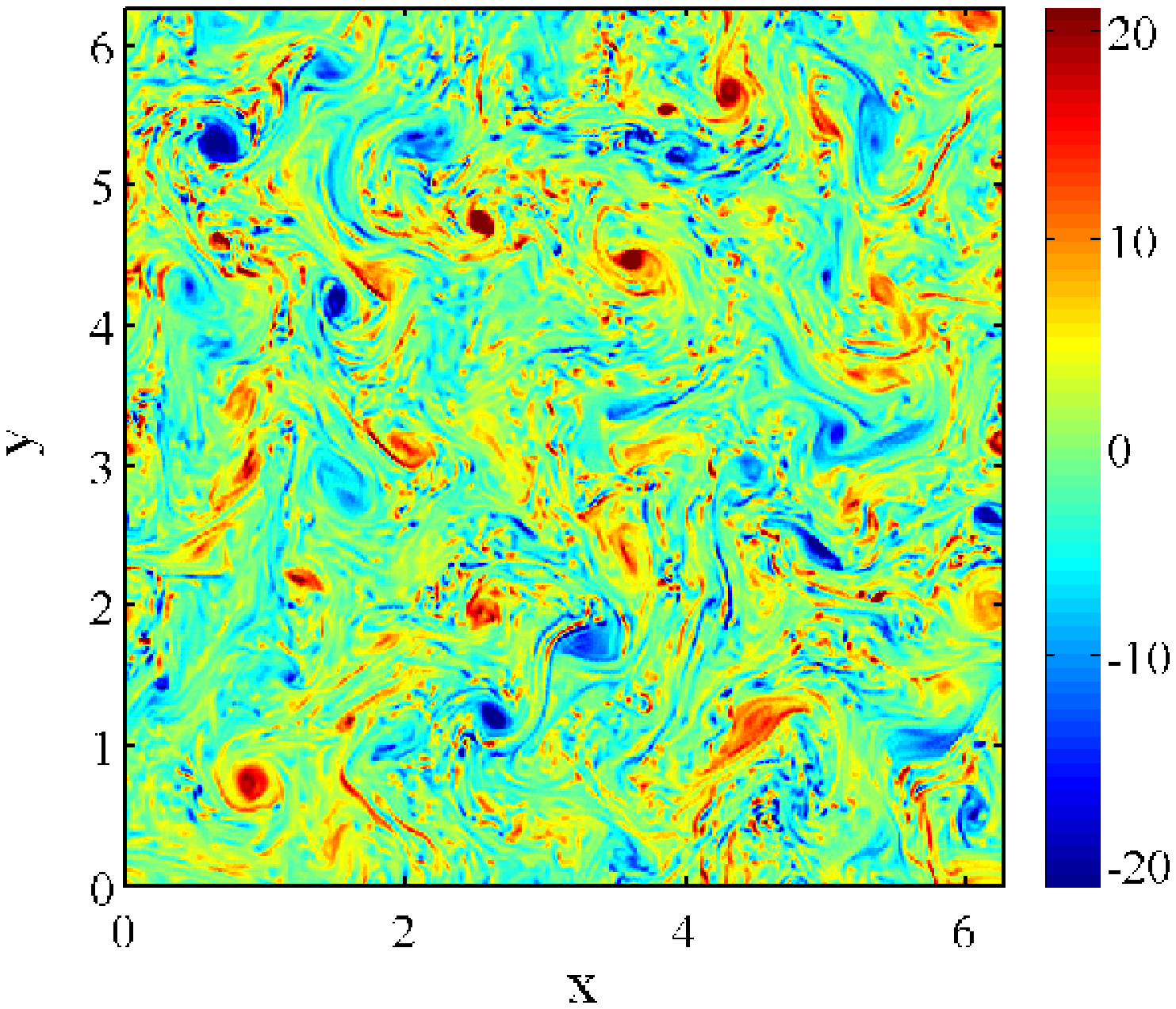}
\includegraphics[width=1.\columnwidth]{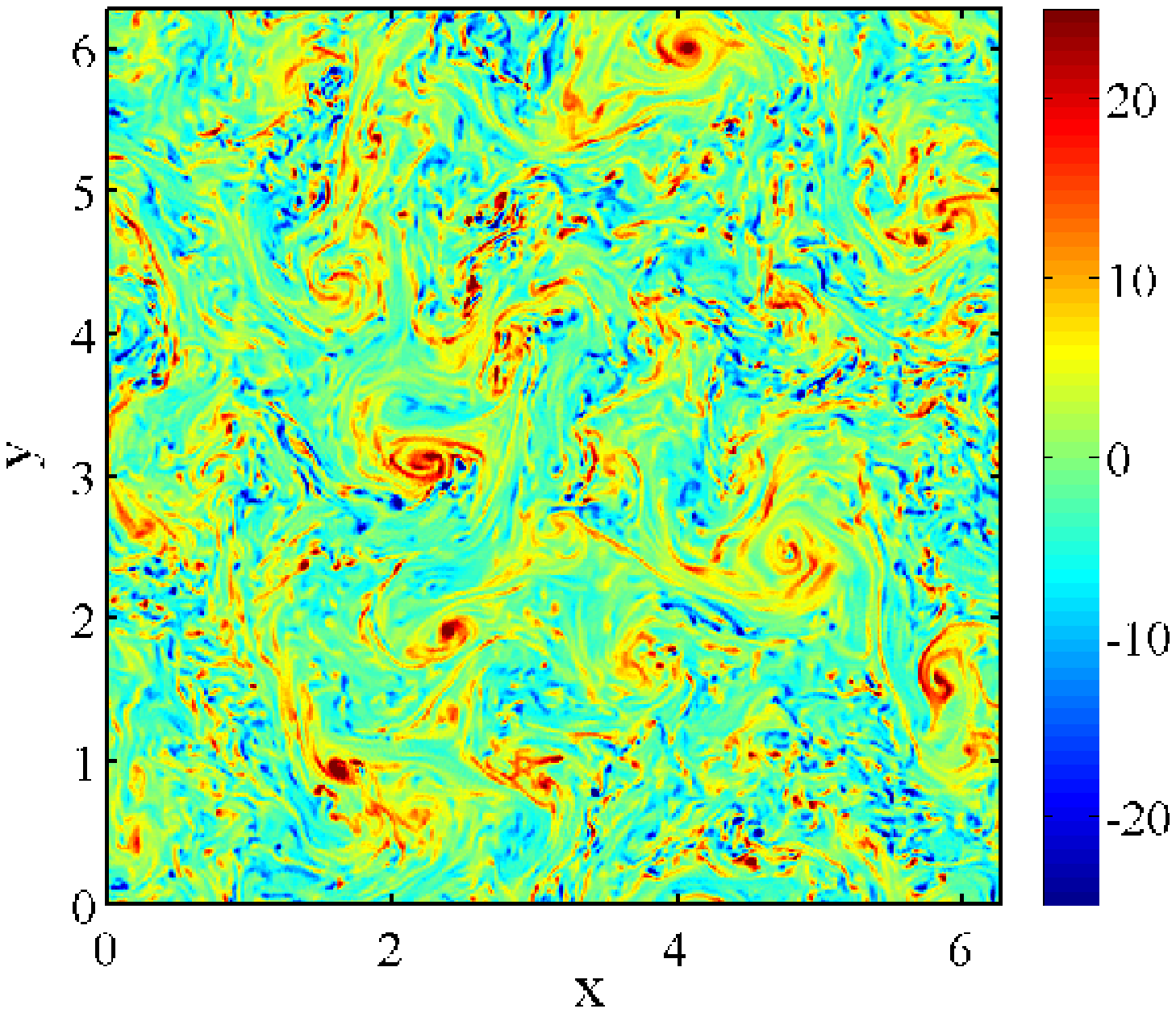}
\caption{Horizontal cuts of vertical vorticity $\omega_z$
for two flow configurations 
$(R=0, S=0.188)$ (left) and $(R=1.5, S=4)$ (right).}
\label{fig6}
\end{figure*}

In the non-rotating case we find that the vortex stretching 
(enstrophy production) is completely suppressed for 
$k < k_z \simeq 5.3 k_f$ ($k_z = 2 \pi /L_z$)
and a direct cascade with almost constant flux of enstrophy is observed 
for $k_f < k < k_z$. 
Therefore in a thin fluid layer the enstrophy behaves as a quasi-invariant, 
\emph{i.e.} it is almost conserved by the dynamics at scales lager than $L_z$. 
In analogy with an ideal 2D flow, the partial conservation of enstrophy 
is responsible for the development of the inverse energy cascade. 

It is worth emphasizing that the total enstrophy behaves as a quasi-invariant,
indicating that the development of the inverse energy cascade 
is not caused directly by the two-dimensionalization of the flow, 
but rather by the presence of a second sign-definite conserved quantity which can be related in spectral space to the energy.
Such mechanism is consistent with the previous findings 
reported in \cite{BMT12, BMT13}, 
which showed that an inverse energy cascade develops 
also in homogeneous isotropic $3D$ turbulent flow 
when mirror symmetry is broken such that helicity 
has a well-defined sign at all wave numbers.

In the rotating, thick case we observe that the vortex stretching 
is also suppressed, but this phenomenon occurs on a broad range of scales
and there is no evidence of an inertial range in which enstrophy 
is conserved. 
Our findings seem to indicate that, 
unlike confinement, moderate rotation is not sufficient to
develop a direct cascade of enstrophy with constant flux.  
Nevertheless, the mechanism by which 
rotation enhances the transfer of energy towards large scales 
is similar to the one which is induced by confinement, 
\emph{i.e.} via suppression of the vortex stretching term.  
It is worth noticing that  
the enstrophy production and enstrophy flux of the two cases 
are almost indistinguishable at small scales.

\begin{figure}[htb!]
\includegraphics[width=1.\columnwidth]{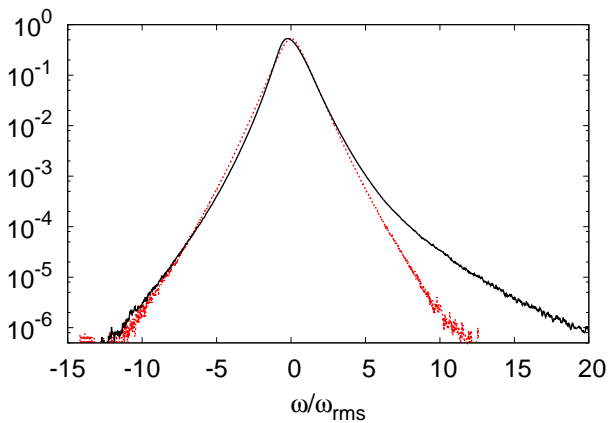}
\caption{PDFs of vertical vorticity for $S=2$, $R=0$ (red dotted line) 
and $S=2$,$R=1.5$ (black solid line). }
\label{fig7}
\end{figure}

\section{Cyclonic-anticyclonic asymmetry}
\label{sec:4}

The breaking of asymmetry between cyclonic and anticyclonic vortices, 
which has been observed both in experiments
\cite{HBG82,HV93,MMR05,SDD08,MMRS11} 
and numerical simulations 
\cite{bartello_metais_lesieur_1994,BB07,BCLGC08}, 
is a distinctive feature of rotating turbulent flows. 
Here we are interested to investigate how this asymmetry is influenced 
both by the rotation and the confinement of the flow. 
In Figure~\ref{fig6} we show two horizontal cuts of the vertical component of 
the vorticity, $\omega_z$, for the two runs discussed in section III,
\emph{i.e.} a thin non-rotating layer and a thick rotating layer with a similar
fraction of energy dissipated at small scales. 
In the non-rotating, thin layer ($R=0$,$S=0.188$, left panel) 
vortices with positive and negative sign are equally distributed.  
In contrast to this, the thick rotating flow ($R=1.5$,$S=4$, right panel)
shows a clear asymmetry between 
cyclonic and anticyclonic vortices. In particular
the cyclonic vortices are enhanced while anticyclonic vortices
are suppressed. 

\begin{figure}[htb!]
\includegraphics[width=1.\columnwidth]{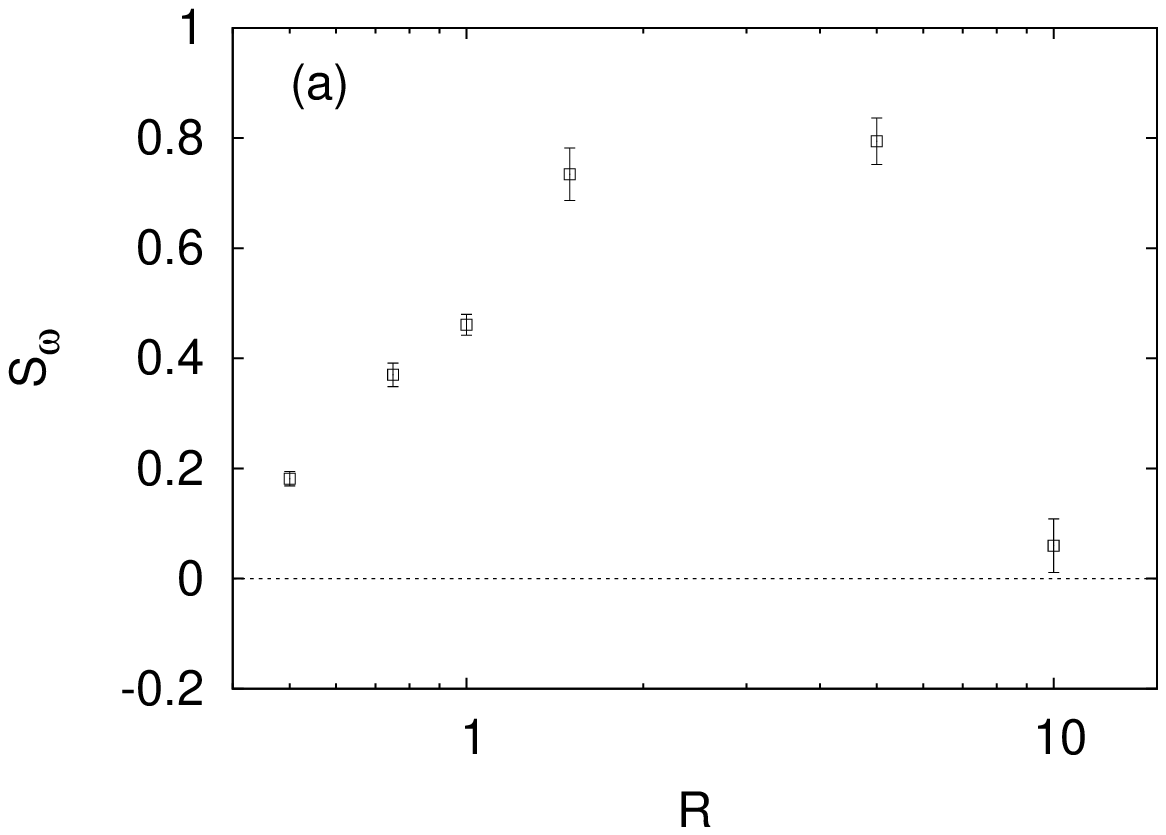}
\includegraphics[width=1.\columnwidth]{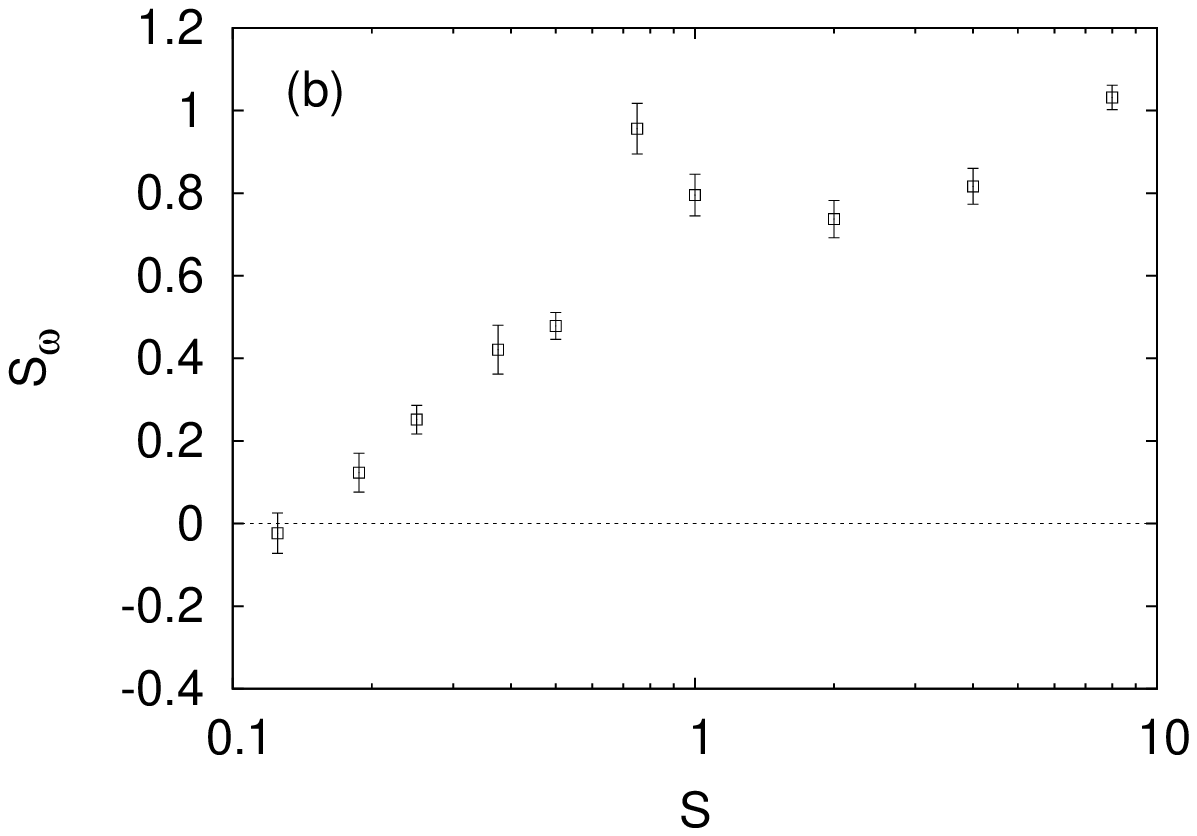}
\caption{
Skewness of vertical vorticity $S_{\omega}$ as a function of (a):
rotation $R$ for fixed $S=2$ and (b): $S$ at fixed $R=1.5$. 
The error bars represent the error of the mean value over a time 
interval.}
\label{fig8}
\end{figure}

In Figure~\ref{fig7} we compare the probability density function (PDF) of
vertical vorticity of the two flows. In the non-rotating case the PDF is
symmetric, while in the rotating case it shows a clear asymmetry. Cyclonic
vortices, corresponding to the right tail, are much more probable than
anti-cyclonic ones. 
We remark that in both cases the mean vertical vorticity 
$\omega_z$ vanishes and therefore a quantitative measure of the asymmetry is 
provided by the skewness 
\begin{equation} 
S_{\omega} = \frac{\langle \omega_z^3 \rangle}
{\langle \omega_z^2 \rangle^{3/2}}.
\label{eq.3.3}
\end{equation}
In the non-rotating case $S_\omega = 0$ while in the rotating cases one has
$S_\omega > 0$. Figure~\ref{fig7} suggest that a
significant contribution to the skewness $S_{\omega}$ comes from the tails of
the PDF of $\omega_z$. 
The $R$-dependence of the skewness is shown in 
Fig.~\ref{fig8}a and it is found to be non-monotonic, with an
increase for moderate rotations followed by a subsequent decrease for larger
rotation rates. This is in agreement with previous findings
\cite{BB07} for which the strongest symmetry
breaking is observed at intermediate $R$, corresponding to $Ro=0.2$. 

It is interesting to point out that the asymmetry also depends 
on the aspect ratio $S$ of the flow.  
Indeed, in the two-dimensional limit 
($S \to 0$) rotation cannot induce an asymmetry in $\omega_z$ as 
the ${\bf \Omega} \times {\bf u}$ term disappears in the 2D version 
of the Navier-Stokes equation (\ref{eq:2.1}).
Therefore, we expect the skewness $S_{\omega}$ to be an increasing
function of $S$ at fixed $R$. 
This is confirmed by our numerical results, as shown in Fig.~\ref{fig8}b. 
Further, our findings suggest that for $S>1$ the skewness 
saturates as $S$ increases, although  
we cannot exclude a different behavior for values of $S$ larger than $8$.
That the asymmetry vanishes for $S \to 0$ also
suggests a possible interpretation of the decreases of the skewness for 
large $R$ as a signature of the bidimensionalization of the flow.

\begin{figure}[htb!]
\includegraphics[width=1.\columnwidth]{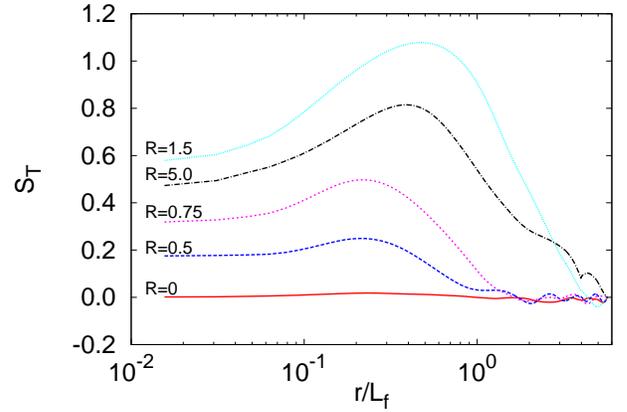}
\caption{Skewness of the transversal velocity increments 
$S_T=\langle \delta u_T^3 \rangle / \langle \delta u_T^{2} \rangle^{3/2}$ 
for $S=2$ and $R=0,0.5,0.75,1.5,5$.}
\label{fig9}
\end{figure}

The asymmetry in the cyclonic-anticyclonic motions does not 
manifests only at small scales but also on large-scale structures, 
as it is evident in Figure~\ref{fig6}. 
In three-dimensional turbulence, 
the leading contributions to the vorticity field
comes from small-scale structures, 
and therefore a criterion based on (\ref{eq.3.3}) 
could be not adequate to capture the asymmetry at large scale. 
Moreover, it is likely to be influenced by the Reynolds number as $\omega_z$ is
a small-scale quantity. 
An alternative measure of the cyclonic-anticyclonic asymmetry 
can be provided by the skewness of the azimuthal velocity increment $\delta u_T$, 
\begin{equation} 
S_{T} = \frac{\langle \delta u_T^3 \rangle}
{\langle \delta u_T^2 \rangle^{3/2}},
\label{eq.3.4}
\end{equation}
where $\delta u_T = \left( \boldsymbol u \left( \boldsymbol x + \boldsymbol r
\right) - \boldsymbol u \left( \boldsymbol x \right) \right) \cdot \boldsymbol
t$, with $\boldsymbol t$ being the horizontal unit vector in the 
cyclonic direction, that is 
$\left( \boldsymbol{r}/r,\boldsymbol{t},\boldsymbol{z} \right)$ 
forms a right-handed system of reference. 
As opposed to (\ref{eq.3.3}), (\ref{eq.3.4}) is a
scale-dependent quantity and it is thus more informative. 
The statistics of azimuthal velocity increment has been fruitfully 
used to analyze 
measurements in the stratosphere \citep{cho_lindborg_2001} and numerical
simulations of rotating and stratified turbulence
\citep{deusebio_augier_lindborg_2013} which show quadratic dependence in $r$. 

In our simulations, we find positive values of $S_T$ of the order of 
unity for rotating flows, (Figure~\ref{fig9}), 
indicating that cyclonic motions are dominating at all scales.
As for the skewness of vorticity, also in this case we observe a non-monotonic
behavior of $S_T$ with $R$ (the maximum value is reached for $R=O(1)$).
Moreover, for all the values $R>0$, we observe a $r$ dependence of the 
skewness which shows a maximum at a scale which increases with $R$.
It is worth noticing that for $R>1$ the asymmetry persists at scales larger
than the forcing scale $L_f$, \emph{i.e.} in the energy inverse-cascade range.

\section{Conclusion and discussion}
\label{sec:5}
In this work we investigated by means of numerical simulations how rotation
affects the turbulent cascade of kinetic energy in a thin fluid layer.  We have
shown that rotation enhances the inverse energy cascade.  This is achieved by a
mechanism which is similar to what is observed when the flow is confined,
\emph{i.e.} by suppressing the production of enstrophy at large scales.  For
thick fluid layers such suppression is sufficient to allow the development of
an inverse cascade even in cases in which it would not be observable in the
absence of rotation.  On the other hand, for a fixed rotation number, we
observe that increasing the aspect ratio causes a suppression of the inverse
energy cascade.
  
Rotation also breaks the symmetry of horizontal flow, 
inducing a predominance of cyclonic vortices over anticyclonic ones. 
Our results confirms that this asymmetry is maximum at 
intermediate rotation rates and vanishes both for weak and strong rotation. 
The analysis of the skewness of transverse velocity structure functions reveals 
that the asymmetry is not only present at small scales, but can be observed
also in large-scale structures of the flow.  Interestingly, we find that the
cyclonic-anticyclonic asymmetry at fixed rotation vanishes as the thickness of
the fluid layer is reduced, consistently with the fact that in ideal
two-dimensional flows the rotation effects disappear.  

When carrying out numerical simulations and experiments in rotation fluids,
great care should thus be taken in setting up the boundary conditions in the
direction parallel to the rotation axis.  In experiments, bottom and top Ekman
boundary layers can strongly influence the dynamics \cite{IT75,HV93,MM06}.
Numerical simulations can similarly be affected by the finite size of
computational domains and confinement effects can effectively influence the
dynamical picture. In rotating flows, bounded domains may also set the minimum
frequency for non-stationary inertial waves, $\omega \sim f k_{z,min}/k$. This
value can nevertheless be very large at large rotation rate, therefore
enforcing a decoupling with the 2D modes which have zero frequency.

Our findings pose interesting questions. 
In particular one may ask whether there would always exist, 
irrespective on how large the rotation rate is, 
a dimensional transition between a 2D dynamics, with an inverse energy cascade, 
and a 3D dynamics, with a direct energy cascade. 
More generally, it would be interesting to derive
theoretical predictions on the scaling of the critical curve in the $(S,R)$
parameter space which borders the region in which the inverse cascade exists. 

\vspace{5mm}
Computer time provided by SNIC (Swedish National Infrastructure for Computing) 
and by Cineca is gratefully acknowledged.
S. Musacchio wish to thank F.~S.~Godeferd for fruitful discussions.  
The authors thank an anonymous reviewer for valuable comments and suggestions.

\bibliography{biblio}

\end{document}